\documentclass[epj,final]{svjour}
\usepackage{epsfig, graphicx}
\usepackage{caption}
\usepackage{subfig}
\begin{document}

\title{Analysis of the exclusive final state npe$^+$e$^-$\\ in quasi-free np reaction }

\author{J.~Adamczewski-Musch$^{4}$, O.~Arnold$^{10,9}$, E.~T.~Atomssa$^{15}$, C.~Behnke$^{8}$, A.~Belounnas$^{15}$, A.~Belyaev$^{7}$, J.C.~Berger-Chen$^{10,9}$, J.~Biernat$^{3}$, A.~Blanco$^{2}$, C.~~Blume$^{8}$
% ...
}

\institute{
\mbox{$^{1}$Institute of Physics, Slovak Academy of Sciences, 84228~Bratislava, Slovakia}\\
\mbox{$^{2}$LIP-Laborat\'{o}rio de Instrumenta\c{c}\~{a}o e F\'{\i}sica Experimental de Part\'{\i}culas , 3004-516~Coimbra, Portugal}\\
\mbox{$^{3}$Smoluchowski Institute of Physics, Jagiellonian University of Cracow, 30-059~Krak\'{o}w, Poland}\\
\mbox{$^{4}$GSI Helmholtzzentrum f\"{u}r Schwerionenforschung GmbH, 64291~Darmstadt, Germany}\\
\mbox{$^{5}$Technische Universit\"{a}t Darmstadt, 64289~Darmstadt, Germany}\\
\mbox{$^{6}$Institut f\"{u}r Strahlenphysik, Helmholtz-Zentrum Dresden-Rossendorf, 01314~Dresden, Germany}\\
\mbox{$^{7}$Joint Institute of Nuclear Research, 141980~Dubna, Russia}\\
\mbox{$^{8}$Institut f\"{u}r Kernphysik, Goethe-Universit\"{a}t, 60438 ~Frankfurt, Germany}\\
\mbox{$^{9}$Excellence Cluster 'Origin and Structure of the Universe' , 85748~Garching, Germany}\\
\mbox{$^{10}$Physik Department E62, Technische Universit\"{a}t M\"{u}nchen, 85748~Garching, Germany}\\
\mbox{$^{11}$II.Physikalisches Institut, Justus Liebig Universit\"{a}t Giessen, 35392~Giessen, Germany}\\
\mbox{$^{12}$Institute for Nuclear Research, Russian Academy of Science, 117312~Moscow, Russia}\\
\mbox{$^{13}$Institute of Theoretical and Experimental Physics, 117218~Moscow, Russia}\\
\mbox{$^{14}$Department of Physics, University of Cyprus, 1678~Nicosia, Cyprus}\\
\mbox{$^{15}$Institut de Physique Nucl\'{e}aire, CNRS-IN2P3, Univ. Paris-Sud, Universit\'{e} Paris-Saclay, F-91406~Orsay Cedex, France}\\
\mbox{$^{16}$Nuclear Physics Institute, The Czech Academy of Sciences, 25068~Rez, Czech Republic}\\
\mbox{$^{17}$LabCAF. F. F\'{\i}sica, Univ. de Santiago de Compostela, 15706~Santiago de Compostela, Spain}\\ 
\mbox{$^{18}$NRC "Kurchatov Institute", PNPI, 188300, Gatchina, Russia}\\
\\
\mbox{$^{a}$ also at ISEC Coimbra, ~Coimbra, Portugal}\\
\mbox{$^{b}$ also at ExtreMe Matter Institute EMMI, 64291~Darmstadt, Germany}\\
\mbox{$^{c}$ also at Technische Universit\"{a}t Dresden, 01062~Dresden, Germany}\\
\mbox{$^{d}$ also at Moscow Engineering Physics Institute (State University), 115409~Moscow, Russia}\\
\mbox{$^{e}$ also at Frederick University, 1036~Nicosia, Cyprus}\\
\mbox{$^{f}$ also at Dipartimento di Fisica and INFN, Universit\`{a} di Torino, 10125~Torino, Italy}\\
\mbox{$^{g}$ also at Utrecht University, 3584 CC~Utrecht, The Netherlands}\\
\mbox{$^{h}$ also at Helmholtz--Institut f\"ur Strahlen-- und Kernphysik, Universit\"at Bonn, Germany}\\
}

\date{Received: date / Revised version: date}

\abstract{We report on the investigation of dielectron production in tagged quasi-free neutron-proton collisions by using a deuteron beam of kinetic energy 1.25~GeV/u inpinging on a liquid hydrogen target. Our measurements with HADES confirm a significant excess of $e^+e^-$ pairs above the $\pi^{0}$ mass in the exclusive channel $dp \to npe^{+}e^{-}(p_{spect})$ as compared to the exclusive channel $ppe^{+}e^{-}$ measured in proton-proton collisions at the same energy. That excess points to different bremsstrahlung production mechanisms. Two models were evaluated for the role of the charged pion exchange between nucleons and double-$\Delta$ excitation combined with intermediate $\rho$-meson production. Differential cross sections as a function of the $e^+e^-$ invariant mass and of the angles of the virtual photon, proton and electrons provide valuable constraints and encourage further investigations on both experimental and theoretical side.
}

\authorrunning{G.~Agakishiev et al.}
\titlerunning{Exclusive $e^{+}e^{-}$ production in the $n-p$ quasi-free reaction}
\maketitle

\begingroup
\renewcommand{\thefootnote}{\alph{footnote}}
\footnotetext{$^{*}$ Corresponding author: witold.przygoda@uj.edu.pl}
\endgroup

\section{Introduction}
\label{intro}

Dielectron production in nucleon-nucleon collisions at kinetic beam energies below the $\eta$ meson threshold production offers a unique possibility to study bremsstrahlung radiation with time-like virtual photons. The relevant final state is $NN\gamma^*(e^+e^-)$ resulting from the interaction between the nucleons or/and their excited states (such as $\Delta$) formed in the collisions. The production amplitude of the virtual photon $\gamma^*$ depends on the electromagnetic structure of the nucleons and on the excited baryon resonances. In the kinematic region of small positive (time-like) values of the squared four-momentum transfer $q^2$ ($q^2>0$), these electromagnetic amplitudes are related to off-shell light vector meson production \cite{mosel}. In general, the bremsstrahlung yield is given by a coherent sum of two types of amplitudes originating from "pure" nucleon-nucleon interactions and intermediate resonance excitation processes. The nucleon contribution provides information on the elastic time-like electromagnetic form factors in a region of four-momentum transfer squared $0<q^2 \ll 4m_p^2$, where $m_p$ is the proton mass, which is inaccessible to measurements in $e^+e^-$ or $\bar{p} p$ annihilation. The resonance contribution includes the production of baryon resonance ($N^*,\Delta$) states. One might visualize this contribution as resonance excitation subsequently decaying into $Ne^+e^-$ via the Dalitz process (since momentum-space diagrams have no time ordering, also other resonance - $Ne^+e^-$ vertices are to be accounted for). This process gives access to the time-like electromagnetic form factors of baryonic transitions in a complementary way to meson photo- or electro-production experiments where negative (i.e. space-like) values of $q^2$ are probed.

Full quantum mechanics calculations have been performed for $np\rightarrow npe^+e^-$ based on effective model Lagrangians \cite{schafer,deJong,kaptari,shyam}, composing the nucleon-nucleon interaction via the exchange of mesons ($\pi,\rho,\omega,\sigma$,..). The virtual photon production happens at $\gamma^* NN$, $\gamma^* NN^{\star}$ and $\gamma^* N\Delta$ vertices and off meson exchange lines. In the energy range relevant for our study, the bremsstrahlung production in proton-proton collisions is dominated by the $\Delta$ resonance excitation. In neutron-proton collisions, however, the nucleon-nucleon contribution plays also a significant role being much stronger (factor 5-10) than in proton-proton collisions. The results of various calculations show some sensitivity to the electromagnetic form factors and to details of the implementation of gauge invariance in the calculations, in particular those related to the emission off the charged pion exchange (for details see discussion in \cite{shyam}). The adjustment of various effects on coupling constants is crucial, too. Consequently, the model cross sections can differ between the models substantially (up to a factor 2-4) in some phase space regions and need to be constrained further by experimental data.    

Another approach, often used in microscopic transport model calculations to account for the nucleon-nucleon bremsstrahlung, is the soft photon approximation \cite{soft,soft1}. It assumes photon emission following elastic nucleon-nucleon interactions with an appropriate phase space modification induced by the produced virtual photon; any interference processes are neglected.  Contributions from the $\Delta$ isobar and higher resonances are added incoherently and treated as separate source of pairs.   

Data on inclusive $e^+e^-$ production in $p-p$, $d-p$ \cite{dls,hades_np} and the quasi-free $n-p$ \cite{hades_np} collisions have been provided by DLS (beam kinetic energy $T=1.04, 1.25$ GeV/u) and HADES ($T=1.25$ GeV/u) Collaborations. The $p-p$ data are well described by calculations with effective Lagrangian models, except \cite{kaptari} which overestimates the measured yields. Various transport models \cite{gibuu,hsd,fuchs}, adding incoherently contributions from $\Delta$ Dalitz decay and from $p-p$ bremsstrahlung (calculated in the soft photon approximation) describe the data well. The dominant contribution is the $\Delta$ Dalitz decay with the dielectron invariant mass distribution slightly depending on the choice of the corresponding transition form-factors \cite{pena,pena2016}.
 
On the other hand, the $d-p$ and particularly the quasi-free $n-p$ data show a much stronger dielectron yield as compared to $p-p$ collisions at the same collision energy. While the yield at the low invariant masses $M_{e^+e^-}<M_{\pi^0}$ could be understood by the larger cross section (by a factor 2) for the $\pi^0$ production in $n-p$ collisions, the differential cross section above the pion mass was underestimated by most of the above mentioned calculations \cite{hades_np}. Even the calculations of \cite{kaptari}, predicting a larger (by a factor $2-4$) bremsstrahlung contribution, fall too short to explain the data in the high mass region. Moreover, it has been demonstrated \cite{hades_np} that a properly scaled superposition of the $p-p$ and $n-p$ inclusive spectra explains dielectron invariant mass distributions measured in $C+C$ collisions at similar energies resolving, from an experimental point of view, the long standing "DLS puzzle" but moving its solution to the understanding of the production in $n-p$ collisions.       
   
Recently, two alternative descriptions have been suggested to explain the enhanced dielectron production in the $npe^+e^-$ final state. The first calculation by Shyam and Mosel \cite{shyam2} is based on the earlier results obtained within the One-Boson Exchange model \cite{shyam} which have been extended to include in the nucleon diagrams the electromagnetic form factors based on the Vector Dominance Model (VDM) \cite{sakurai}. The results show a significant improvement in the description of the inclusive data,  mainly due to the effect of the pion electromagnetic form-factor in the emission of $e^+e^-$ from a charged exchange pion. Its presence enhances the dielectron yield at large invariant masses. Such a contribution can also be interpreted as a formation of a $\rho$-like final state via annihilation of the exchanged charged pion with a pion from the nucleon meson cloud. Since the charged pion exchange can only contribute to the $np\rightarrow npe^+e^-$ final state but not to the $pp\rightarrow ppe^+e^-$ (note that this is valid only for the exclusive final states) it explains in a natural way the observed difference between the two reactions.

The second calculation by Bashkanov and Clement \cite{clement} also addresses a unique character of the $n-p$ reaction for a production of the $\rho$-like final state via the charged current. Here the mechanism of the $\rho$ production is different and proceeds via the interaction between two $\Delta$s created simultaneously by the excitation of the two nucleons. Indeed, such a double-$\Delta$ excitation is known to be an important channel for the two-pion production at these energies \cite{hades_2pi,wasa} and is governed by the t- or u-channel meson exchange.  The amplitude for the transition of the $n-p$ system to the $NN\rho$ final state via a $\Delta-\Delta$ state is proportional to the respective isospin recoupling coefficients ($9j$-symbols) which for the $p-p$ reactions is zero. 

It is important to stress that all aforementioned calculations were performed for the exclusive $npe^+e^-$ final state whereas the experimental data were analysed in the inclusive $e^+e^-X$ channels. The comparisons were not direct, since other channels, besides the exclusive $npe^+e^-$ channel, can also contribute. For example, the $\eta$ Dalitz decay in the $d-p$ collisions has to be considered in calculations due to the finite nucleon momentum distribution inside the deuteron providing an energy in the $np$ reference frame above the meson production threshold. Various calculations show, however, that the inclusion of this channel is not sufficient for the full description of the data. Moreover, also other channels, like the $np\rightarrow de^+e^-$ proposed in \cite{martem} or  bremsstrahlung radiation accompanied by one or two pions in the final state can contribute to the inclusive production as well.

The main goal of investigating the exclusive reaction $np \rightarrow npe^+e^-$ is two-fold: (i) to verify whether the observed enhancement of the inclusive dielectron production over $p-p$ data has its origin in the exclusive final state and (ii) to provide various multi-particle differential distributions of the exclusive final state to characterize the production mechanism and provide more constraints for the comparison to models. 

Our work is organized as follows. In Section \ref{exp} we present experimental conditions, apparatus and principles of the particle identification and reconstruction. We also explain the method of  selection of the exclusive channel and the normalization procedure. In Section \ref{sim} we discuss our simulation chain composed of the event generator, modelling of the detector acceptance and the reconstruction efficiency. In Section \ref{results} we present various differential distributions characterizing the $npe^+e^-$ final state and compare them to model predictions, followed by the conclusions and outlook in Section~\ref{summary_outlook}.

\section{Experiment and data analysis}
\label{exp}
\subsection{Detector overview}

The High Acceptance Dielectron Spectrometer (HADES) consists of six identical sectors placed between coils of a superconducting magnet instrumented with various tracking and particle identification detectors. The fiducial volume of the spectrometer covers almost the full range of azimuthal angles and polar angles from  $18^\circ$- $85^\circ$  with respect to the beam axis. The momentum vectors of produced particles are reconstructed by means of the four Multiwire Drift Chambers (MDC) placed before (two) and behind (two) the magnetic field region. The experimental momentum resolution typically amounts to $2-3\%$ for protons and $1-2\%$ for electrons, depending on the momentum and the polar emission angle. Particle identification (electron/pion/kaon/proton) is  provided by a hadron blind Ring Imaging Cherenkov (RICH) detector, centred around the target, two time-of-flight walls based on plastic scintillators covering polar angles $\theta>45^\circ$ (TOF) and $\theta<45^\circ$ (TOFino), respectively, and a Pre-Shower detector placed behind the TOFino. The magnetic spectrometer is associated at the forward region ($0.5^\circ-7^\circ)$ by a high granularity Forward Wall (FW) placed 7 meters downstream of the target. The Forward Wall consists of 320 plastic scintillators arranged in a matrix with cells of varying sizes and time resolution of about 0.6 ns. In particular, it was used for identification of the spectator proton from the deuteron break-up. 

A detailed description of the spectrometer, track reconstruction and particle identification methods can be found in \cite{hadesspec}.

In the experiment a deuteron beam with a kinetic energy of $T=1.25$ GeV/u and intensities of up to $10^7$ particles/s was impinging on a $5$ cm long liquid-hydrogen target with a total thickness of $\rho d=0.35$ g/cm$^2$. The events with dielectron candidates were selected by a two-stage hardware trigger: (i) the first-level trigger (LVL1) demanding hit multiplicity $\geq 2$ in the TOF/TOFino scintillators, in coincidence with a hit in the Forward Wall detector; (ii) the second-level trigger (LVL2) for electron identification requiring at least one ring in the RICH correlated with a fast particle hit in the TOF or an electromagnetic cascade in the Pre-Shower detector \cite{hadesspec}.

\subsection{Normalization}
\label{elastic}

The normalization of experimental yields is based on the quasi-free proton-proton elastic scattering measured in the reaction $d+p \rightarrow pn p_{spect}$ within the HADES acceptance ($\theta_{CM}^p \in (46^\circ-134^\circ)$). The known cross section of the $p-p$ elastic scattering has been provided by the EDDA experiment \cite{edda}. The events were selected using a dedicated hardware trigger requesting two hits in the opposite TOF/TOFino sectors. The proton elastic scattering was identified using conditions defined on (a) two-track co-planarity $\Delta \phi=180^\circ \pm 5^\circ$ and (b) the proton polar emission angles $tan (\theta_1)\times tan(\theta_2)=1/\gamma^2_{CM}= 0.596\pm 0.05$. These constraints account for the detector resolution and the momentum spread of the proton bound initially in the deuteron. The latter one was simulated using realistic momentum distributions implemented in the PLUTO event generator \cite{pluto}. The measured yield was corrected for the detection and the reconstruction inefficiencies and losses in the HADES acceptance due to the incomplete azimuthal coverage. The overall normalization error (including the cross section deduced from the EDDA data) was estimated to be $7\%$ \cite{hades_2pi}.


\subsection{Acceptance and reconstruction efficiency}
\label{eff}

To facilitate the comparison of the data with the various reaction models the geometrical acceptance of the HADES spectrometer has been computed and tabulated as three-dimensional matrices depending on the momentum, the polar and the azimuthal emission angles for each particle species ($p$, $e^{+}$, $e^{-}$). The resolution effects are modelled by means of smearing functions acting on the generated momentum vectors (the matrices and smearing functions are available upon request from the authors).  

The efficiency correction factors were calculated individually as one-dimensional functions of all presented distributions. The calculations were performed using a full analysis chain consisting of three steps: (i) generation of events in the full space according to a specific reaction model, described in Section \ref{sim}, (ii) processing of the events through the realistic detector acceptance using the GEANT package and (iii) applying specific detector efficiencies and the reconstruction steps as for the real data case. The respective correction functions are calculated as ratios of the distributions obtained after steps (ii) and (iii).    

In Section \ref{results} we also present various angular distributions corrected for the detector acceptance. Those correction factors were calculated as two-dimensional functions of the dielectron invariant mass and the given angle using two reaction models (described in details in Sec.~\ref{sim}). The difference between both models were used to estimate systematic errors related to model corrections. The models were verified to describe the measured distributions within the HADES acceptance reasonably well. For those cases we also present original distributions measured inside the acceptance.

\subsection{Selection of the npe$^+$e$^-$ final state}
\label{channel_selection}

\begin{figure*}

  \resizebox{1.0\textwidth}{0.3\textheight}{
    \includegraphics[width=1.0\textwidth]{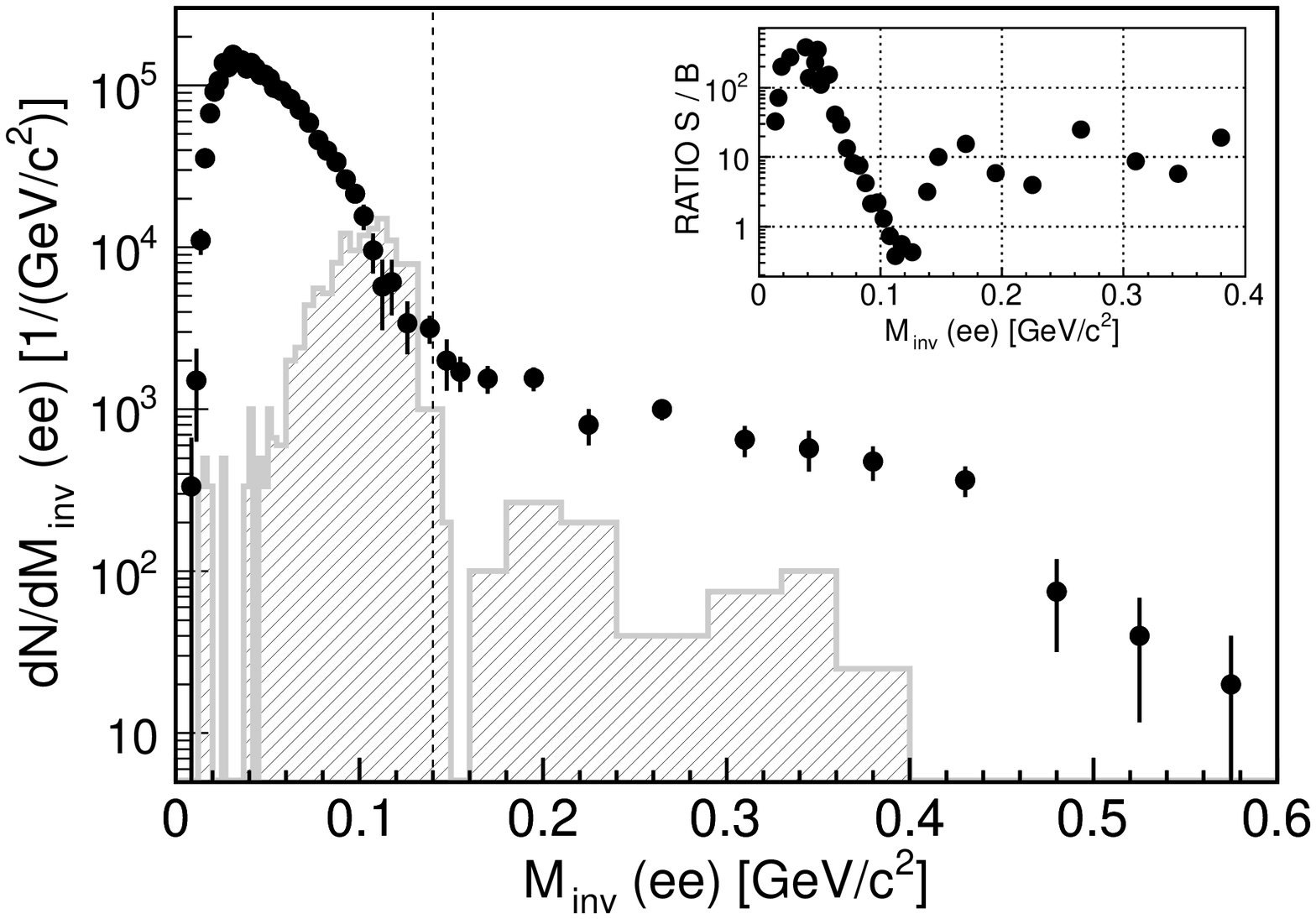}
    \includegraphics[width=1.0\textwidth]{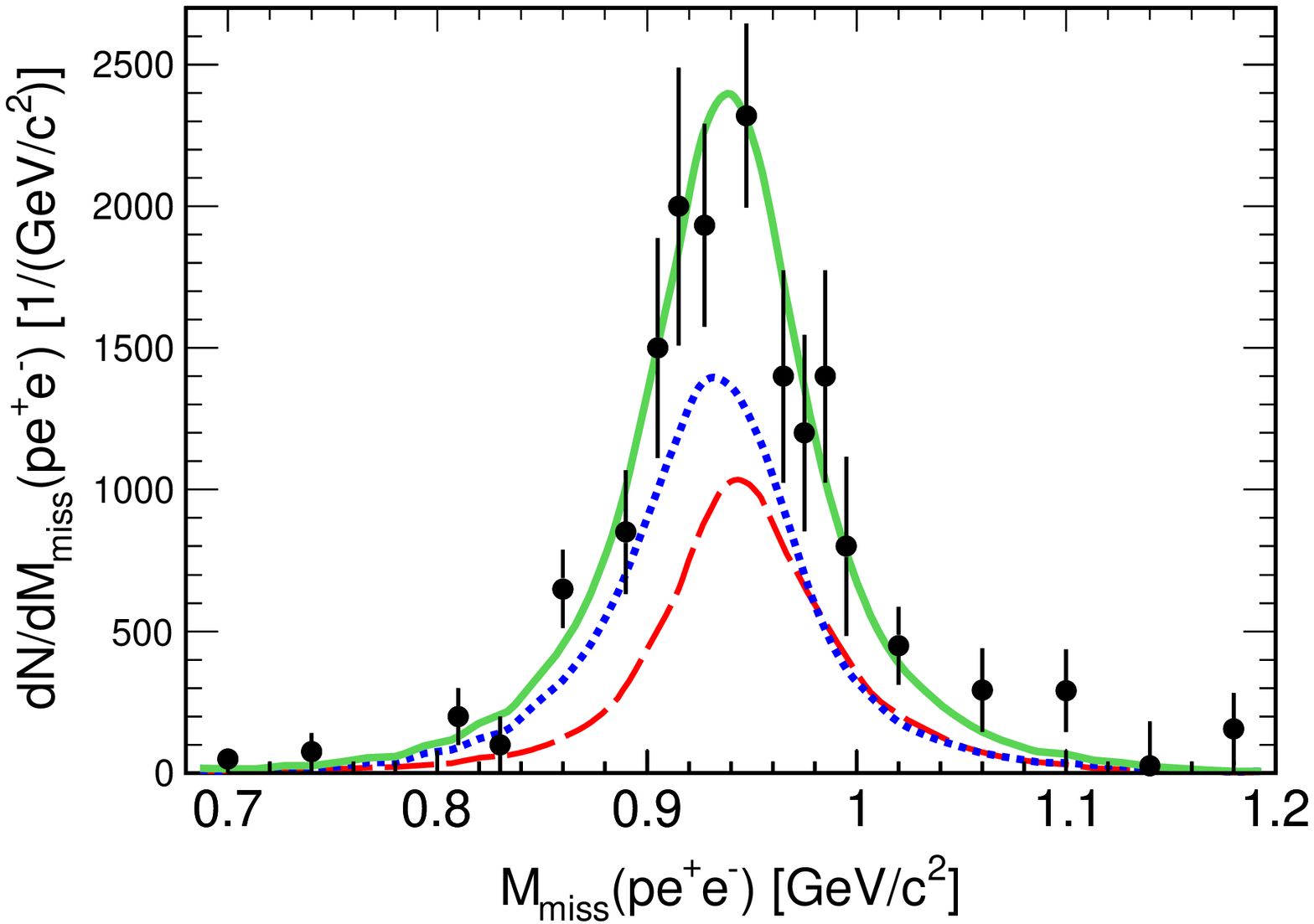}
  }

\caption{(Color online) Events with $npe^+e^-$ final state. Left: invariant mass distributions of $e^+e^-$ signal pairs (black dots), the combinatorial background (CB) (hatched histogram) and the signal/background ratio (inset). Experimental data (black dots) are within the HADES acceptance and not corrected for reconstruction inefficiency. Right: the $pe^+e^-$ missing mass for $np \to pe^+e^- X$ reaction and dielectron invariant masses $M_{e+e-}>0.14$ GeV/c$^2$ (dots) overlayed with a Monte-Carlo simulation (green curve) normalized to the same yield as the data. Two major contributions of model A are depicted: dotted blue curve - $\rho$-meson contribution, dashed red curve - $\Delta$ contribution (see text for details). In both cases, the number of counts is given per GeV/c$^2$ to account for the variable bin width. Only statistical errors are indicated.}
\label{ee_raw}   
\end{figure*}

The procedure of identification of the $npe^+e^-$ final state is initiated by the event selection requesting (i) at least one track with a positive charge, (ii) at least one dielectron pair (like-sign or unlike-sign) detected in the HADES, and (iii) at least one hit in the FW. The electron and positron tracks are identified by means of the RICH detector, providing also emission angles for matching the rings with tracks reconstructed in the MDC, and the time of flight difference of the tracks measured by the TOF/TOFino detectors. Proton identification is achieved by a two-dimensional selection on the velocity ($\beta=v/c$) and the momentum reconstructed in the TOF/TOFino detectors and the tracking system, respectively. There was no dedicated start detector in our experiment, therefore, the reaction time was calculated from the time-of-flight of the identified electron track. The spectator proton was identified as the fastest hit in the FW within the time of flight window of 5 ns spanned around the central value of 26 ns expected for the proton from the deuteron break-up. Such a broad window takes into account both the detector resolution ($\pm 4\sigma$) and the much smaller effect of the spectator momentum distribution (about $\pm 8\sigma$).  

Further, for all $pe^+e^-$ candidates in an event, the missing mass for $np\to p e^+e^- X$ was calculated, assuming the incident neutron carrying half of the deuteron momentum. The exclusive $npe^+e^-$ final state was finally selected via a one-dimensional hard cut centred around the mass of the neutron $0.8 < M_{pe^+e^-}^{miss} < 1.08$ GeV/c$^2$. A variation of this selection has no influence on the data at $M_{inv}(e^+e^-) >$ 0.14 GeV/c$^2$ and introduces a systematic error on the yield of about 10\% for the $\pi^{0}$ region, as deduced from comparisons to Monte Carlo simulations.

The same procedure was also applied for the $pe^-e^-$ and the $pe^+e^+$ track combinations in order to estimate the combinatorial background (CB) originating mainly from a multi-pion production followed by a photon conversion in the detector material. The CB was estimated, using the like-sign pair technique, calculated for every event with a proton: $dN_{CB}/dM = 2\sqrt{(dN/dM)_{++} (dN/dM)_{--}}$. The signal pairs are obtained by the CB subtraction: $dN^{e^+e^-}_{SIG}/dM = dN^{e^+e^-}_{ALL}/dM - dN_{CB}/dM$.

The resulting $e^+e^-$ invariant mass distributions of the signal and the CB are shown in Fig. \ref{ee_raw} (left panel) together with the signal to background ratio (inset) for the identified $pe^+e^-$ events. In the invariant mass region above the prominent $\pi^0$ Dalitz decay peak, the signal is measured with a small background. In Fig. \ref{ee_raw} (right panel), the missing mass distribution of the $pe^+e^-$ system with respect to the projectile-target is shown for the events with the invariant mass $M_{e^+e^-}>0.14 $ GeV/c$^2$. The data are compared to a Monte Carlo simulation - green solid curve (model A, see Section~\ref{sim} for details). Its total yield has been normalized to the experimental yield to demonstrate the very good description of the shape of the distribution. One should note that a slight shift of the peak position (0.944 GeV/c$^2$) and, particularly, a broadening of the missing mass distribution ($\sigma=0.037$ GeV/c$^2$) is caused by the momentum distribution of the neutron in a deuteron, which is accounted for in the simulation. The spectrometer resolution causes half of the measured width.

\section{Comparison to models: event generation and simulation}
\label{sim}

The most recent calculations of Shyam and Mosel \cite{shyam2} and Bashkanov and Clement \cite{clement} offer an explanation of inclusive dielectron data measured in $n-p$ collisions at $T=1.25$ GeV. A characteristic feature of both models is an enhancement in the dielectron invariant mass spectrum for $M_{e^+e^-}>0.3$ GeV/c$^2$ due to the intermediate $\rho$-like state in the in-flight emission by the exchanged charged pions, which are present in the case of the $np \to npe^+e^-$ reaction, unlike in the $pp\to ppe^+e^-$ reaction. A major difference between the models is that the charged pions are exchanged between two $\Delta$s in \cite{clement} and between two nucleons in \cite{shyam2}. We have chosen these models as a basis for our simulation (described in details below). 

The model \cite{clement} assumes a sub-threshold $\rho$-meson production, via intermediate double delta $\Delta^{+} \Delta^{0}$ or $\Delta^{++} \Delta^{-}$ excitation, and its subsequent $e^+e^-$ decay, according to a strict Vector Dominance Model (VDM) \cite{sakurai}. The total cross section, for the  $np\rightarrow \Delta\Delta$ channel, has been predicted to be $\sigma_{\Delta\Delta}=170~\mu b$. Events generated with the theoretical differential distributions and characterized by the $np$ and the $\gamma^{\star}$ four-vectors, have been provided by the authors \cite{clement_priv}. The dielectron decays of the $\gamma^*$ have been modeled in our simulations following the VDM prescription for the $\rho$-meson differential decay rate (see \cite{clement}) and assuming the isotropic electron decay in the virtual photon rest frame. 

The remaining dielectron sources ($\pi^0$, $\Delta$ and $\eta$ Dalitz decays) were computed using the PLUTO event generator. The detailed description of the procedure was published in \cite{hades_np,pluto}, and in fact the calculations in \cite{clement} use exactly the same method. For the $\Delta$ Dalitz decay, the "QED model" was used, with the constant electromagnetic Transition Form Factors (eTFF) fixed to their values at the real-photon point. As a consequence, the Coulomb form factor is neglected and the $e^+$ or $e^-$ angular distribution with respect to the $\gamma^*$ in the rest frame of the $\gamma^*$, is taken as $\propto 1+\cos^2 \theta$, in agreement with data \cite{witek}. 

The channels included in our simulations are the following ones: (i) $np \rightarrow \Delta^{+,0}(n,p)\rightarrow np \pi^0 \rightarrow np e^+e^-\gamma$  (ii)  $np\rightarrow np\eta\rightarrow np e^+e^- \gamma$ and (iii) $np \rightarrow \Delta^{+,0} (n,p) \rightarrow (n,p) e^+e^- (n,p)$. One should note that the latter channel accounts for the part of the bremsstrahlung radiation related to the $\Delta$ excitation, since the pre-emission graphs associated with the $\Delta$ excitation have a small contribution \cite{shyam}. We assume that one-pion production is dominated by the $\Delta$ excitation which saturates the $I=1$ component of the $n-p$ reaction. The iso-scalar component of the $n-p$ reaction at our energy is much smaller, as shown by \cite{andrej,bystr}, and has been neglected. The cross section $\sigma_{\Delta^{+,0}}$ for the production of the $\Delta^+$ and $\Delta^0$ resonances in the $n-p$ reactions has been deduced in \cite{teis} within the framework of the isobar model by a fit to the available data on one-pion production in nucleon-nucleon reactions and amounts to $\sigma_{\Delta^+}=\sigma_{\Delta^{0}}=5.7 $ mb. Furthermore, in the simulation we have included angular distributions for the production of the $\Delta$ excitation deduced from the partial wave analysis of the one-pion production in the $p-p$ collisions at the same energy \cite{hades_pwa}. These distributions provide a small correction with respect to the one-pion exchange model \cite{teis}, which were originally included in the PLUTO generator.

The contribution of the $\eta$ (see \cite{pluto} for details of the implementation) to the exclusive $npe^+e^-$ channel is negligible but was included for comparison with the calculations of the inclusive production \cite{hades_np}, where it plays an important role. This model is later referred as the model A.

The model of Shyam and Mosel \cite{shyam2} is based on a coherent sum of $NN$ bremsstrahlung and isobar contributions. It demonstrates a significant enhancement of the radiation in the high-mass region due to contributions from the charged internal pion line and the inclusion of the respective electromagnetic pion form factor. This mechanism modifies the contribution of the bremsstrahlung radiation from the nucleon charge-exchange graphs, which, as pointed out in the introduction, are absent in the case of the $pp \to ppe^+e^-$ reaction. The other part of the bremsstrahlung corresponds to the $\Delta$ excitation on one of the two nucleon lines and its subsequent Dalitz decay ($Ne^+e^-$). Although the latter dominates the total cross section at $M_{e^+e^-}<0.3$ GeV/c$^2$, the modified nucleon-nucleon  contribution makes a strong effect at higher masses. Unfortunately, the proposed model does not provide details about angular distributions of the final state particles.  In our simulation we use the bremsstrahlung generator included in the PLUTO package \cite{pluto} with the respective modification of the dielectron invariant mass distribution to account for the results of \cite{shyam2}. Since there is no guidance in the model on angular distributions of the protons and of the virtual photons, we have assumed the distribution introduced in the model A for the $\Delta$ production. We denote this model as the model B. 

The modeling of the quasi-free $np$ collisions has been implemented in both models based on a spectator model \cite{pluto}. This model assumes that only one of the nucleons (in our case the neutron) takes part in the reaction while the other one, the proton, does not interact with the projectile and is on its mass shell. The momenta of the nucleons in the deuteron rest frame are anti-parallel and generated from the known distribution \cite{benz}.

\section{Results}
\label{results}

\begin{figure}
\hspace{0mm}
\resizebox{0.5\textwidth}{0.4\textheight}{ \includegraphics{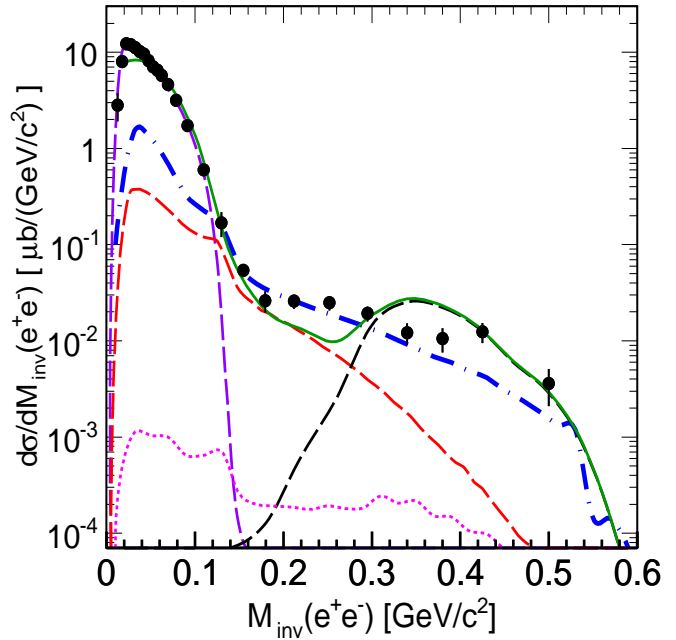}}
\caption{(Color online) Dielectron differential cross section as a function of the invariant mass of $e^+e^-$ within the HADES acceptance. The data (black dots) are corrected for the detection and reconstruction inefficiency and presented per GeV/c$^2$. The simulated cocktail (curves) of the $\pi^0$ (dashed violet), $\eta$ (dotted magenta), $\Delta$ (dashed red) Dalitz decays, $\rho$ from the double $\Delta-\Delta$ interaction process (dashed black) according to the model \cite{clement} and the sum (contributions from $\pi^{0}$, $\eta$, $\Delta$ and $\rho$ - solid green curve) - model A. The dotted-dashed blue curve shows the bremsstrahlung contribution from \cite{shyam} - model B.}

\label{eep_th}       
\end{figure}

The exclusive final state $np\gamma^*$ can be characterized by five independent variables selected in an arbitrary way. Assuming azimuthal symmetry in the production mechanism, only four variables are needed. The decay of the $\gamma^*$ into the $e^+e^-$ pair can be characterized by two additional variables. In this work we have chosen the following observables: 

(i) the three invariant masses of the $e^+e^-$ pair ($M_{e^+e^-}$, equivalent to the $\gamma^*$ mass), the proton-$e^+e^-$ system ($M_{pe^+e^-}$) and of the proton-neutron ($M_{np}$) system, respectively

ii) the two polar angles of the proton ($cos^{CM}(\theta_p)$) and of the virtual photon ($cos^{CM}(\theta_\gamma^*)$) defined in the center-of-mass system and the polar angle of the lepton (electron or positron) in the $\gamma^*$ rest frame ($cos(\theta^{e-\gamma^*}_{\gamma^*})$) with respect to the direction of the $\gamma^*$ in the c.m.s.

In the next sections we present the corresponding distributions and compare them to the results of our simulations. The experimental distributions are corrected for the reconstruction inefficiencies (see paragraph \ref{eff}) and are presented as differential cross sections within the HADES acceptance, after normalization, as described in paragraph \ref{elastic}. We present also acceptance corrected angular distributions.

\subsection{Invariant mass distributions}

The dielectron invariant mass distributions is very sensitive to the coupling of the virtual photon to the $\rho$-meson. Therefore we start the presentation of our data with Fig.~\ref{eep_th} which displays the dielectron invariant mass distribution and a comparison to the simulated spectra. As already observed in the case of the inclusive $e^+e^-$ production \cite{hades_np}, the  $e^+e^-$  yield in the $\pi^0$ region is found to be in a very good agreement with the $\pi^0$ production cross section  of 7.6 mb used as an input to the simulation (see Sec.~\ref{sim}). One should note that the contribution from $np\rightarrow np\pi^0 (\pi^0\rightarrow e^+e^-\gamma$) channel could not be completely eliminated by the selection on the $pe^+e^-$ missing mass (paragraph~\ref{channel_selection}) due to the finite detector mass resolution. This contribution is well described by our simulations, confirming the assumed cross section of the one-pion production. The good description obtained in the exclusive case  demonstrates in addition that the  acceptance on the detected proton and the resolution of the $pe^+e^-$ missing mass are well under control. 

The distribution for invariant masses larger than the $\pi^0$ mass ($M_{e^+e^-}>M_{\pi^0}$) is dominated by the exclusive $np\to npe^+e^-$ reaction (as also proven by the missing mass distribution in Fig.~\ref{ee_raw} - right panel), which is of main interest for this study. In this mass region the general features of the dielectron yield are reproduced by the model A. The $\Delta$ Dalitz decay dominates for the $e^+e^-$ invariant mass between 0.14 GeV/c$^2$ and 0.28 GeV/c$^2$, while the $\rho$ contribution prevails at higher invariant masses. The $\eta$ Dalitz decay  gives a negligible contribution. A closer inspection reveals that the $\Delta$ Dalitz alone cannot describe the yield in the mass region  $0.14<M_{e^+e^-}<0.28$ GeV/c$^2$. This is not surprising since the nucleon-nucleon bremsstrahlung is also expected to contribute in this region. On the other hand, the $\rho$ contribution overshoots the measured yield at higher masses, even in a stronger way, than observed in the case of the inclusive data \cite{clement}. The low mass cut of the $\rho$ contribution is due to the threshold at the double-pion mass, which should be absent in the case of the dielectron decay but is the feature of the applied decay model \cite{clement}.

\begin{figure*}
  \resizebox{1.0\textwidth}{0.3\textheight}{
    \includegraphics[width=1.0\textwidth]{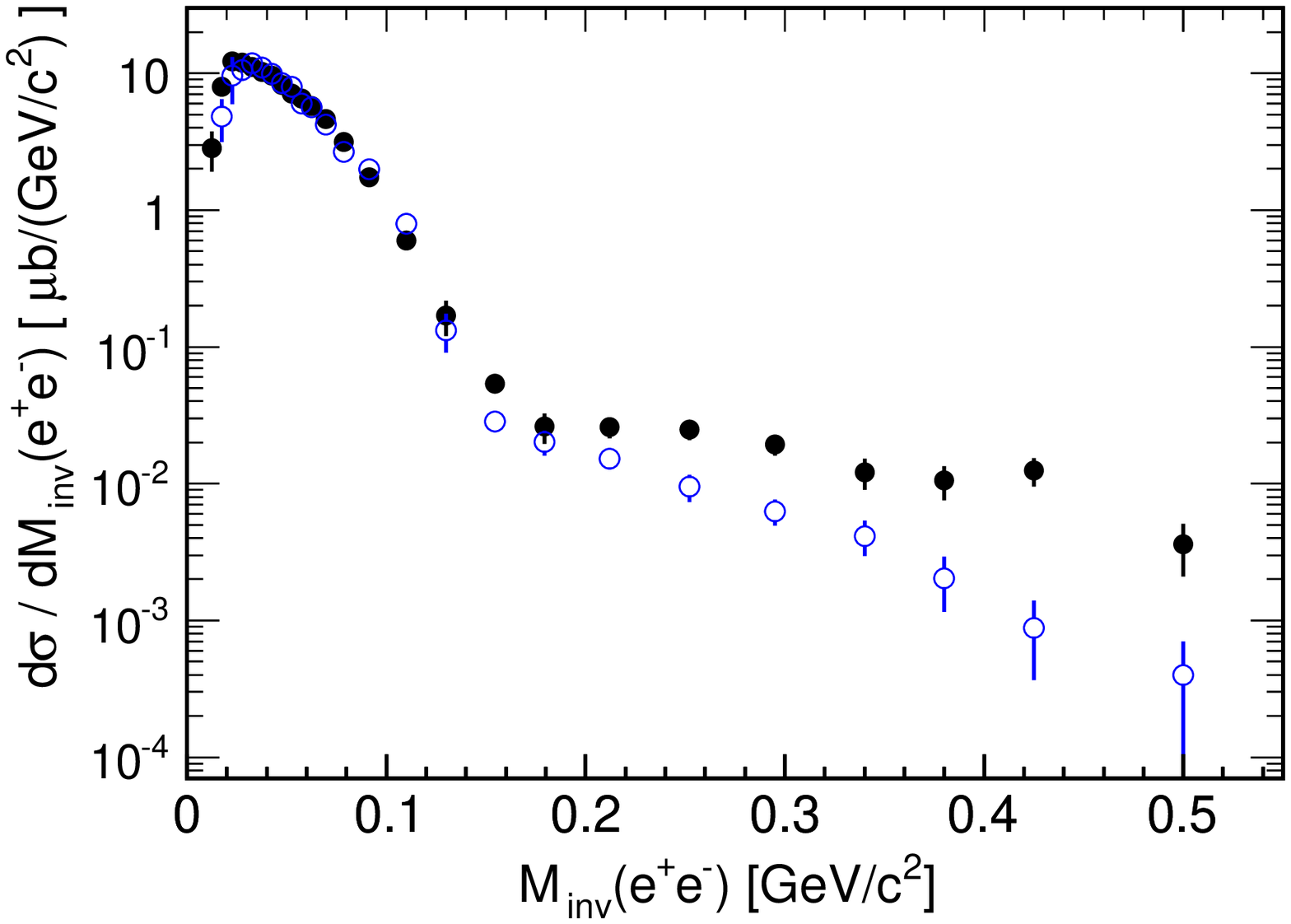}
    \includegraphics[width=1.0\textwidth]{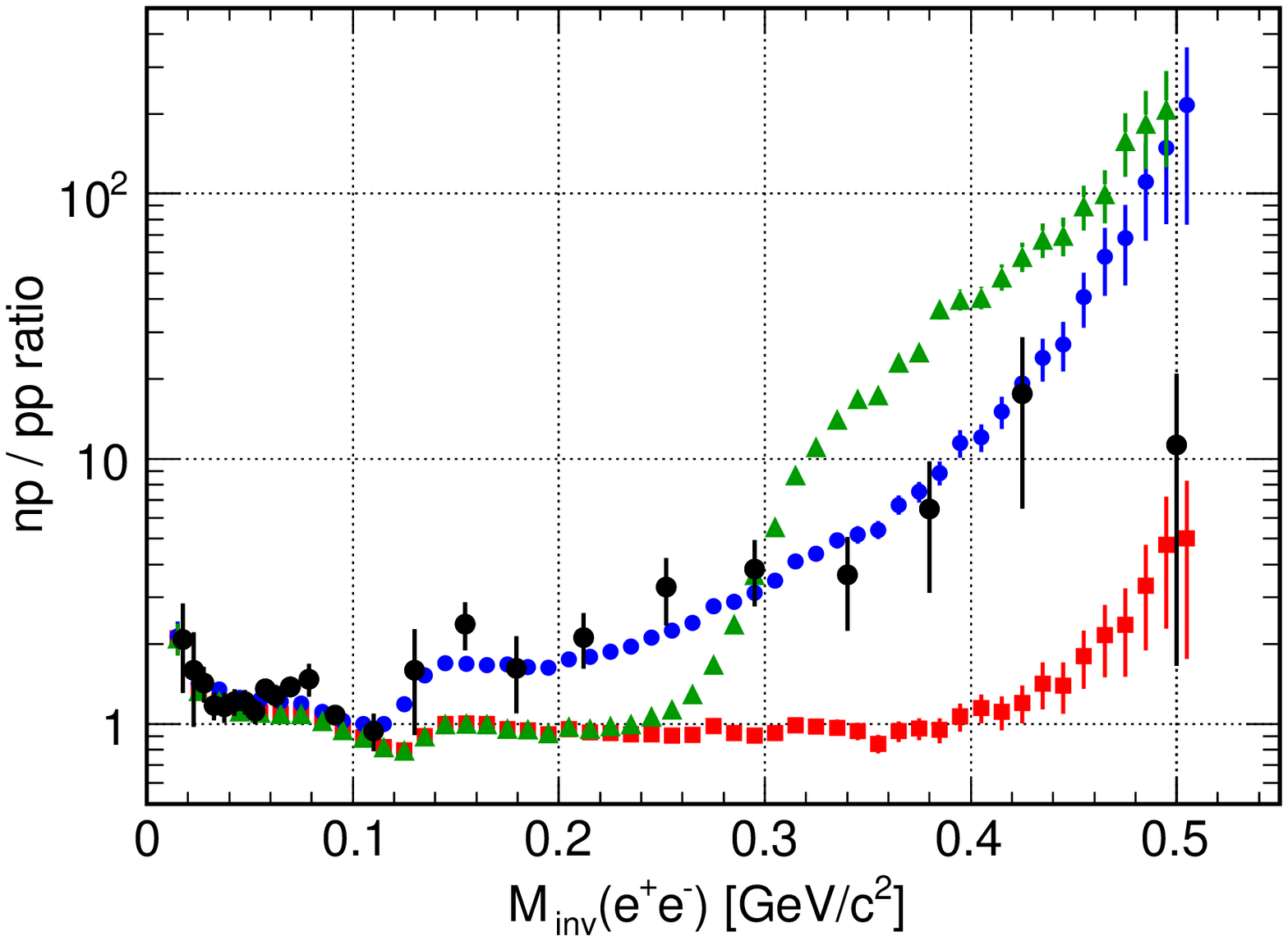}
  }

\caption{(Color online) Left: dielectron invariant mass distribution for $npe^+e^-$ (black dots) and $ppe^+e^-$ (blue open circles) normalized to the same $\pi^0$ cross section within the HADES acceptance. Right: the ratio of the differential cross sections (in absolute scale, within the HADES acceptance) from $np \to npe^+e^-$ and $pp \to ppe^+e^-$ exclusive channels (black dots). The ratio of the model (A and B) and the $p-p$ Monte-Carlo simulation is presented by green triangles (model A) and blue dots (model B). In addition a difference in phase volumes between $n-p$ and $p-p$ collisions in the aforementioned channels is estimated (red squares). For details see text.
}
\label{ee_pp}       
\end{figure*}

The simulation based on the model B presents a rather different shape, with a smooth decrease of the yield as a function of the invariant mass. It was indeed shown \cite{shyam} that the introduction of the pion electromagnetic form factor at the charged pion line enhances significantly the yield above the $\pi^0$ peak, but does not produce any structure. The yield for $M_{e^+e^-}< $ 0.14 GeV/c$^2$ is strongly underestimated, which is expected, due to the absence of $\pi^0$ Dalitz process in the model, which aimed only at a description of the $np\to np e^+e^-$. Above the $\pi^0$ peak,  model B comes in overall closer to the data than model A, but it underestimates the yield at the very end of the spectrum ($M_{e^+e^-} > $ 0.35 GeV/c$^2$). The exclusive yield calculated within the model B might slightly depend on the hypothesis we have made on the angular distributions (see paragraph~\ref{sim}). The expected effect is however rather small, since the proton angular distribution is well described by the simulation, as will be shown in paragraph~\ref{ang}. The comparison of the simulations based on both models to the experimental dilepton invariant mass distributions seem to favour the explanation of the dielectron excess due to  the electromagnetic form factor on the charged pion line, as suggested in \cite{shyam2}. 

The exclusive invariant mass distribution can be also compared with the $ppe^+e^-$ final state measured by the HADES at the same beam energy \cite{witek}. The latter one is well described, as discussed in Section~\ref{intro}, by various independent calculations which all show the dominance of the $\Delta$ Dalitz decay process for invariant masses larger than 0.14 GeV/c$^2$. Thus, it can serve as a reference for the identification of some additional contributions appearing solely in the $npe^+e^-$ final state. Figure~\ref{ee_pp} (left panel) shows the comparison of the $e^+e^-$ invariant mass distributions normalized to the $\pi^0$ production measured in the reaction $np\rightarrow npe^+e^-$. It reveals a different shape above the pion mass. 

The right panel of Fig.~\ref{ee_pp} shows the ratio of both differential cross sections, with their absolute normalization, as a function of the invariant mass in comparison to three different simulations. The error bars plotted for data and simulations are statistical only. First, we note that the ratio of the two cross sections in the $\pi^0$ region within the HADES acceptance and inside the $M_{pe^+e^-}$ missing mass window amounts to $\sigma_{\pi^0}^{np}/\sigma_{\pi^0}^{pp}=1.48 \pm 0.24$, which is well reproduced by the simulations for the $\pi^0$ Dalitz decay. The ratio of the cross sections in the full solid angle is 2, according to the measured data \cite{hades_pwa} and as expected from the isospin coefficients for the dominant $\Delta$ contribution. However, the ratio measured inside the HADES acceptance is smaller because it is reduced by the larger probability to detect a proton in addition to the $e^+e^-$ pair for the $ppe^+e^-$ final state as compared to $npe^+e^-$. For the $e^+e^-$ invariant masses larger than the pion mass, the ratio clearly demonstrates an excess of the dielectron yield in the exclusive $n-p$ channel over the one measured in $p-p$. It indicates an additional production process which is absent in the $p-p$ reactions, as proposed by the discussed models. 

In order to exclude trivial effects, like the different phase space volumes available in the $p-p$ and quasi-free $n-p$ collisions due to the neutron momentum spread in the deuteron, first we plot the ratio of the cross sections of $\Delta$ channels in both reactions (red squares on the right panel of Fig.~\ref{ee_pp}). An enhancement is indeed present but only at the limits of the available phase space. It confirms that the phase space volume difference gives a very small contribution to the measured enhancement in the $npe^+e^-$ channel. 

The green triangles (model A) and blue dots (model B) in Fig.~\ref{ee_pp} (right panel) represent the ratio of the respective model simulation and the $p-p$ Monte Carlo simulation: the sum of $\pi^0$ and $\Delta$ Dalitz decays ($\Delta$ with a point-like eTFF) \cite{witek}. The ratios take into account the differences in the phase volume between $n-p$ and $p-p$, as mentioned above. Similar to the comparison of the dielectron invariant mass distribution in Fig.~\ref{eep_th}, the calculation of \cite{shyam2} (model B) gives a better description of the data for the invariant masses larger than the $\pi^0$ mass. 

\begin{figure}

\hspace{0mm}
\resizebox{0.48\textwidth}{0.25\textheight}{
\includegraphics{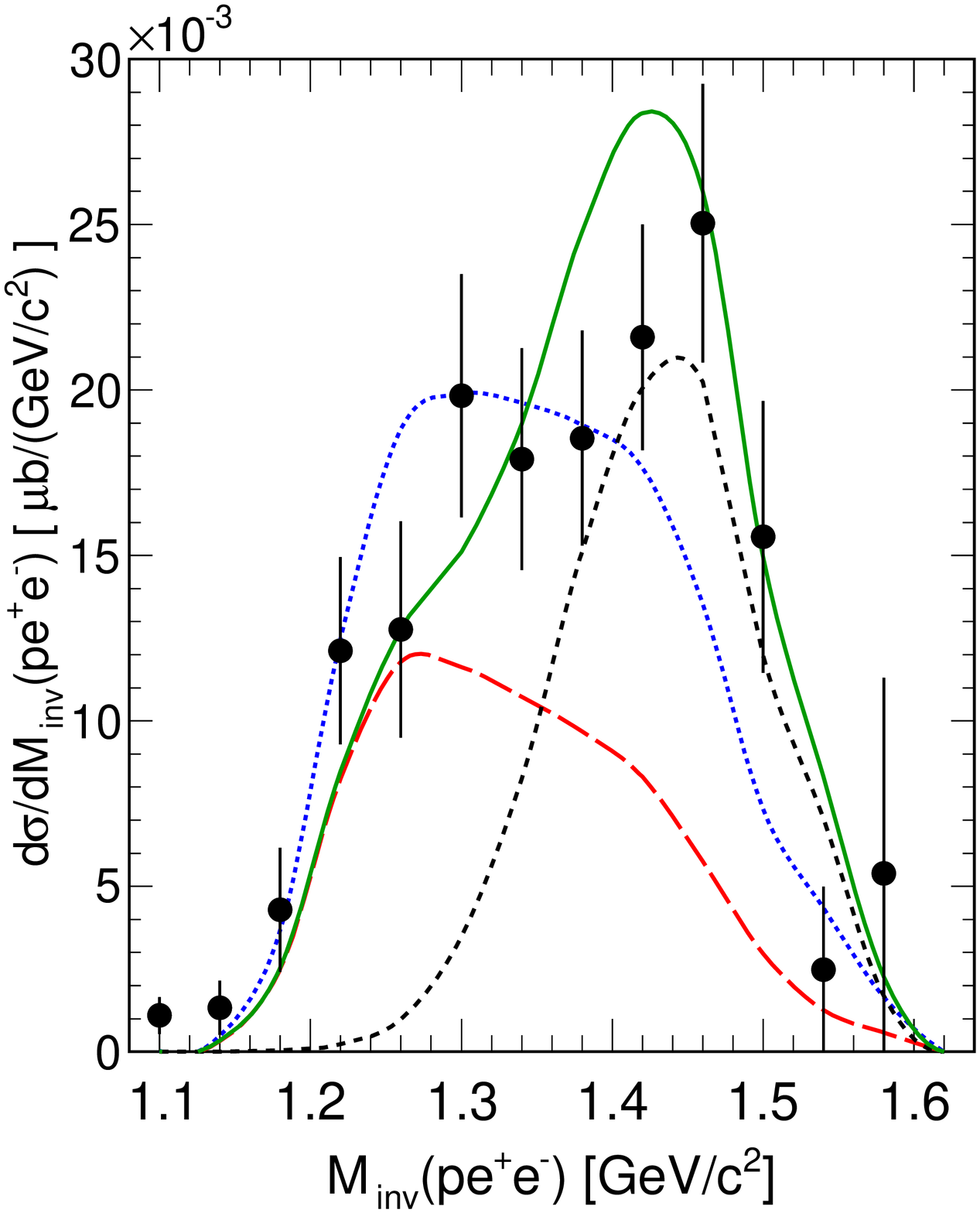}\includegraphics{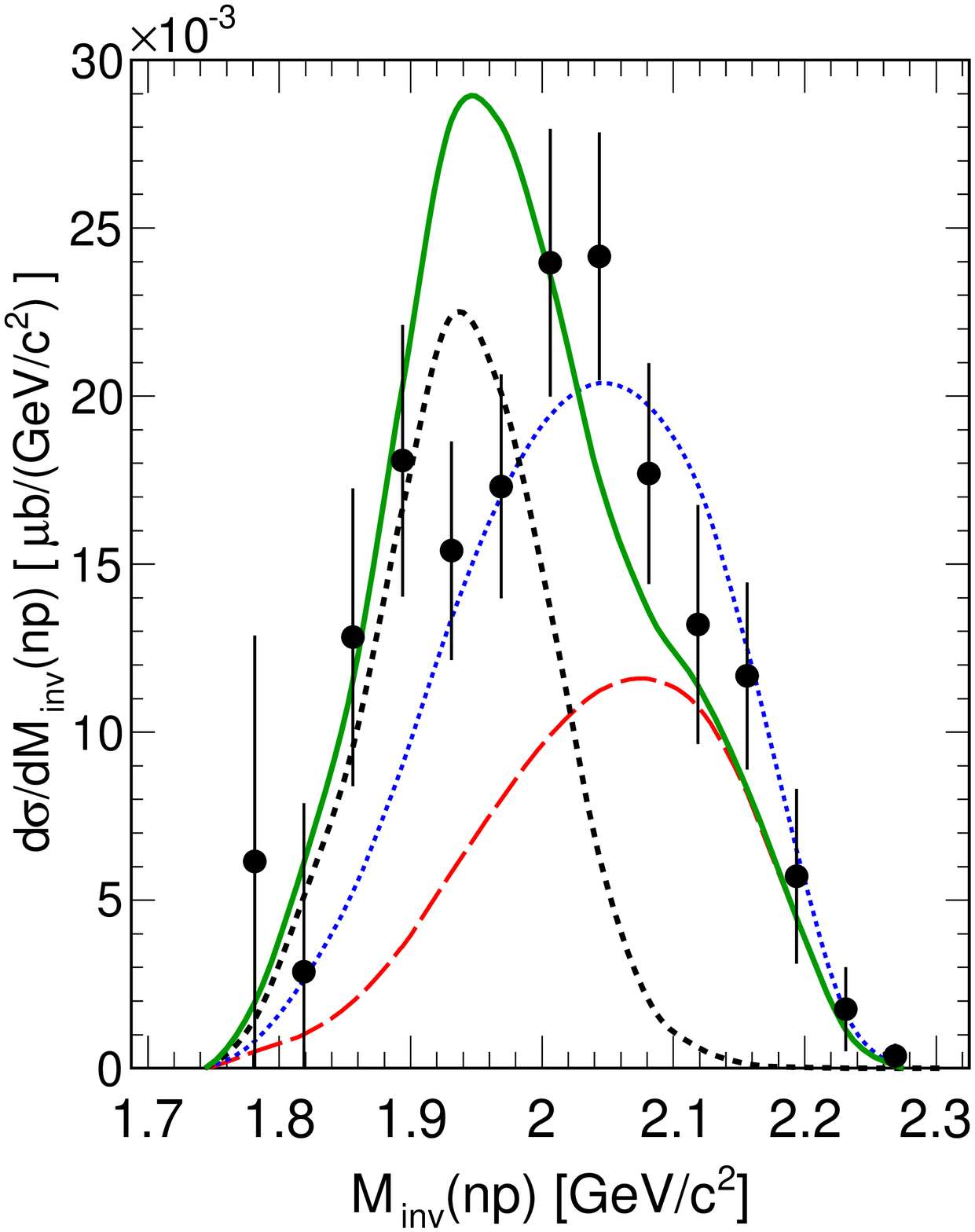}}

\caption{(Color online) $npe^+e^-$ final state within the HADES acceptance. Differential cross section as a function of the invariant mass of $pe^+e^-$ (left panel) and of $np$ (right panel) compared to model A (solid green), with the following components: $\Delta$ Dalitz (dashed red), $\rho$-meson decay from the double-$\Delta$ interaction (short-dashed black) and model B (blue dotted).}

\label{eep1_th}       
\end{figure}

Figure~\ref{eep1_th} shows the two other invariant mass distributions of the $pe^+e^-$ ($M_{pe^+e^-}$, left panel) and the $np$ ($M_{np}$, right panel) systems. Both distributions are plotted for masses of the virtual photon $M_{e^+e^-}>0.14$ GeV/c$^2$ and are compared to the models A and B. For the model A, the $\Delta$ and $\rho$ contributions are shown separately. As expected, the distribution at low $M_{pe^+e^-}$ is dominated by low mass dielectrons, originating mainly from the $\Delta$ decays (we note that the observed shape in the simulation is due to an interplay between $\Delta^+\rightarrow pe^+e^-$ and $\Delta^0 \rightarrow ne^+e^-$ decays, both contributing with same cross sections) and at higher masses by the $\rho$-like channel. On the other hand, the invariant mass distribution of the $np$ system is dominated at low masses by the $\rho$ contribution, which in the model A overshoots slightly the data. In general, the high-mass enhancement visible in the $e^+e^-$ mass spectrum is consistently reflected in the shapes of the two other invariant mass distributions.

\subsection{Angular distributions}
 \label{ang}
In the discussion of the angular distributions we consider separately two bins of the dielectron invariant mass: $0.14<M_{e^+e^-}<0.28$ GeV/c$^2$ and $M_{e^+e^-}>0.28$ GeV/c$^2$. The selection of the two mass bins is dictated by the calculations which suggest two possible different production regimes, with a dominance of  the $\rho$-like contribution in the second bin. 

\begin{figure}

\hspace{0mm}
\resizebox{0.48\textwidth}{0.25\textheight}{
\includegraphics{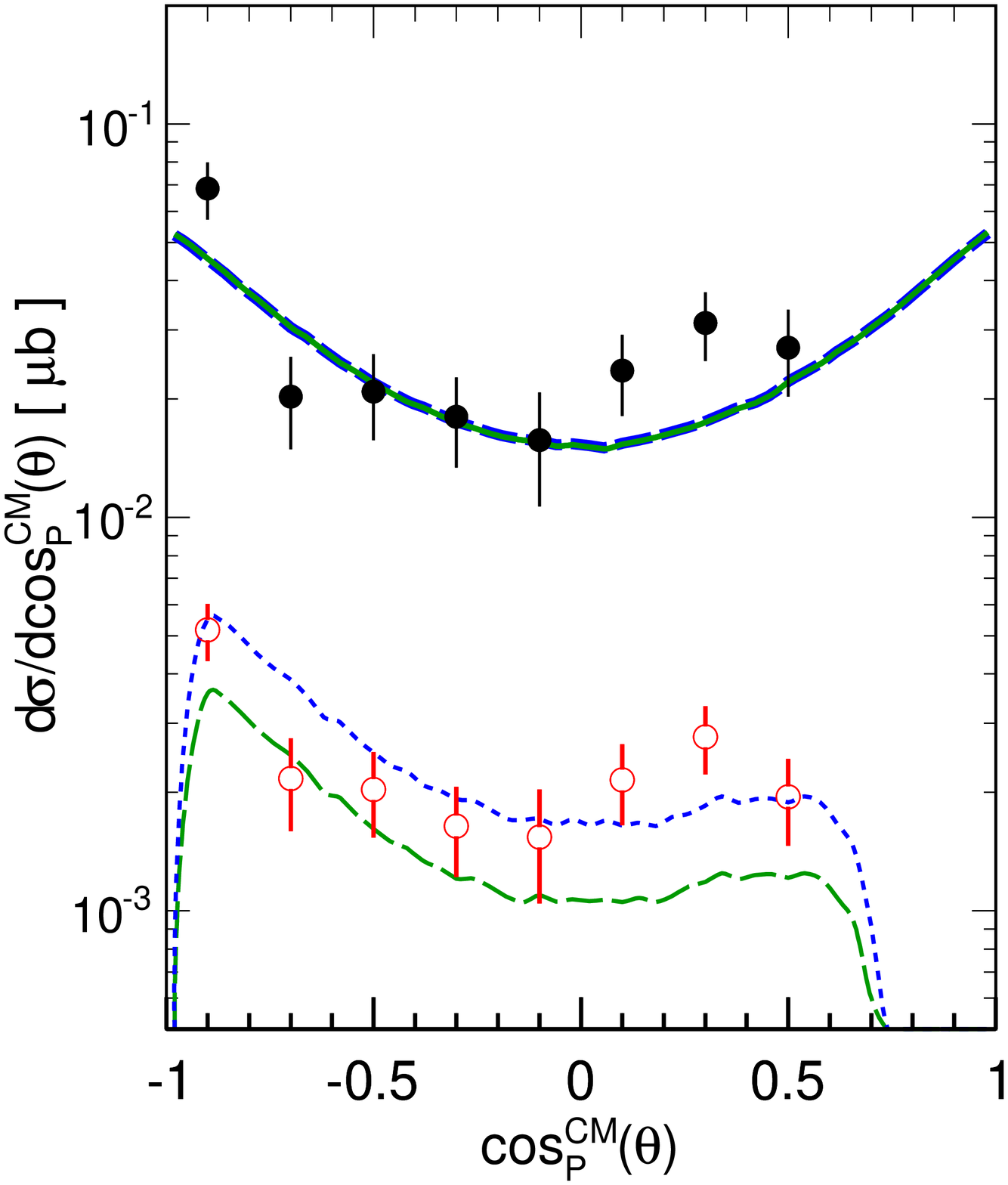} \includegraphics{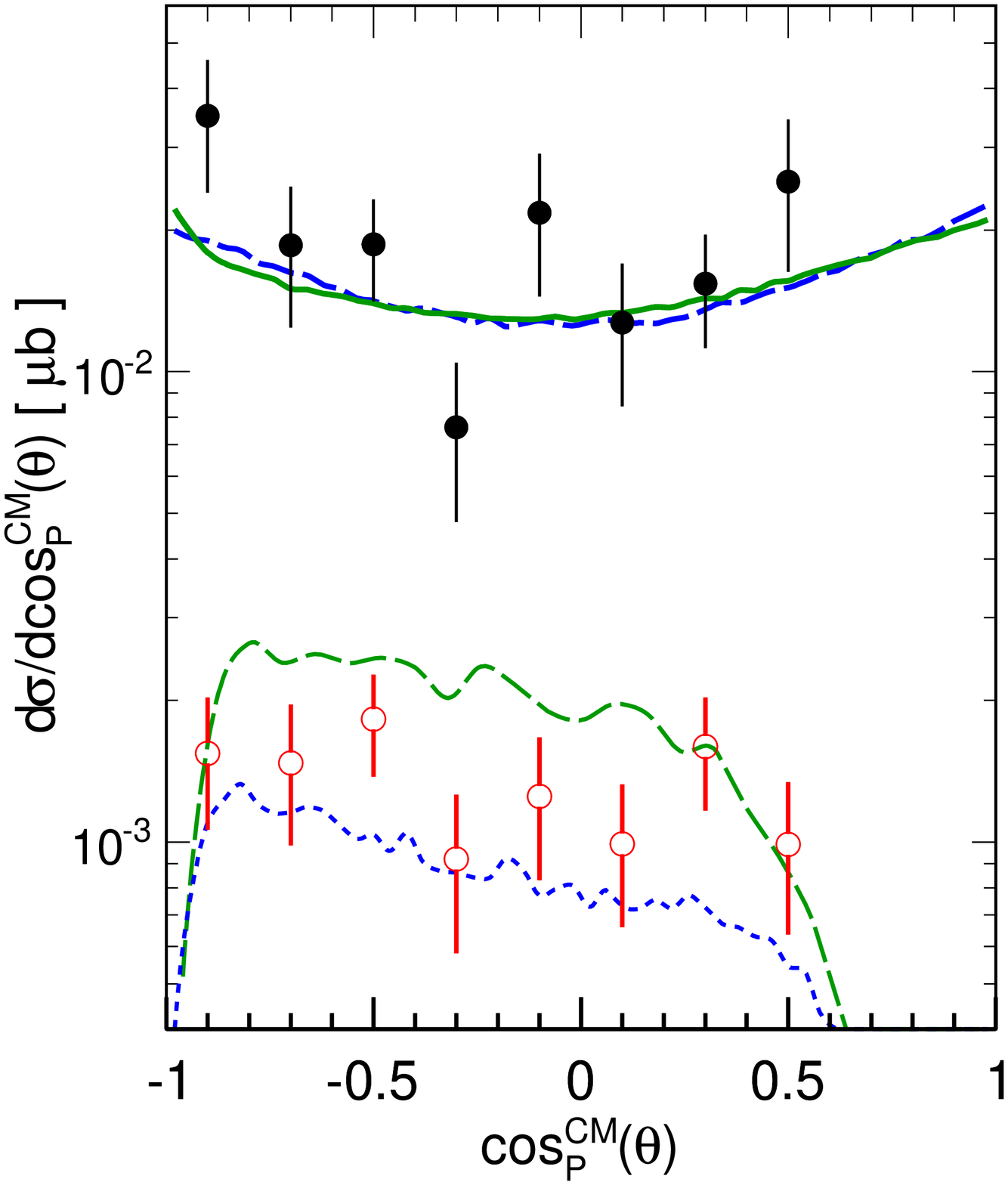}}

\caption{(Color online) Differential cross sections for the dilepton production in $npe^+e^-$ channel as a function of the proton emission angle in the c.m.s.: for $0.14 < M_{e^+e^-} < 0.28$ GeV/c$^2$ (left panel) and $M_{e^+e^-}>0.28$ GeV/c$^2$ within the HADES acceptance (open red dots) and the full solid angle (full black dots). The solid curves display predictions from the simulations in the full solid angle normalized to the experimental yield: the green curve represents model A (in the low mass bin mostly $\Delta$, in the high mass bin mostly $\rho$), dashed blue represents model B. The dotted/dashed curves are within the detector acceptance for model A (dashed green) and B (dotted blue) (see text for details), respectively.}
\label{cos_p}       
\end{figure}

Figure~\ref{cos_p} displays the differential angular distributions of the proton in the c.m.s., both within the HADES acceptance and after acceptance corrections. In the first case, the experimental distributions are compared to the predictions of the simulations on an absolute scale. In the second case, the simulated distributions are normalized to the experimental yield after acceptance corrections in order to compare the shapes. The acceptance correction applied to the data has been calculated as described in paragraph~\ref{eff}. 
 
As can be deduced from Fig.~\ref{eep_th}, according to model A, the low-mass bin is dominated in the simulation by the $\Delta$ Dalitz decay process, while the $\rho$-like contribution determines the dielectron production in the higher mass bin. In the first mass bin, the distribution exhibits a clear anisotropy, pointing to a peripheral mechanism. The simulated distributions for the models A (dashed green curve) and B (dotted blue curve) differ in magnitude but have similar shapes. This is due to the fact that the angular distribution for the model B is the same as in the $\Delta$ contribution of model A, which dominates in this mass region (see Sec.~\ref{sim}) - both contributions have the same angular distribution in the full solid angle (solid green and superimposed dashed blue curves). The shape of the experimental angular distribution is rather well taken into account by both simulations, where the angular distributions for the $\Delta$ production from the PWA analysis is used, leading to a symmetric forward/backward peaking. However, there is an indication for some enhancement above the simulation in the $npe^+e^-$ channel for the forward emitted protons, unfortunately cut at small angles by the HADES acceptance. It might be due to the charge exchange graphs involving nucleons, which are not properly taken into account by the symmetric angular distribution used as an input for the simulation. Indeed, in the case of the $\Delta$ excitation, charge exchange and non-charge exchange graphs have the same weight, which yields a symmetric angular distribution for the proton in the c.m.s.. This is different for nucleon graphs, where the contribution of the charge exchange graphs to the cross section are enhanced due to the isospin coefficients by a factor 4 and, therefore, forward emission of the proton is favoured. 

For the higher invariant $e^+e^-$ masses, the angular distribution is  more isotropic and is described rather well by both simulations which again exhibit similar characteristics. The flattening of the distributions reflects the different momentum transfers involved in the production of heavy virtual photons. However, as already mentioned, the angular distribution in model B follows the $\Delta$ production angular distribution, while in model A it is properly calculated for the $\rho$ production via the double-$\Delta$ mechanism. 

It is interesting to observe that the two angular distributions are very similar. In particular, the distribution with respect to $cos_{p}^{CM}(\theta)$ from the model A is symmetric, although graphs with emission of the neutron from a $\Delta^-$ excited on the incident neutron (and corresponding emission of the proton from the excitation of a $\Delta^{++}$ on the proton at rest) are highly favoured by isospin factors and induce a strong asymmetry for the production of the $\Delta$s, as shown for example in \cite{Huber94}.

\begin{figure}

\hspace{0mm}
\resizebox{0.48\textwidth}{0.25\textheight}{
\includegraphics{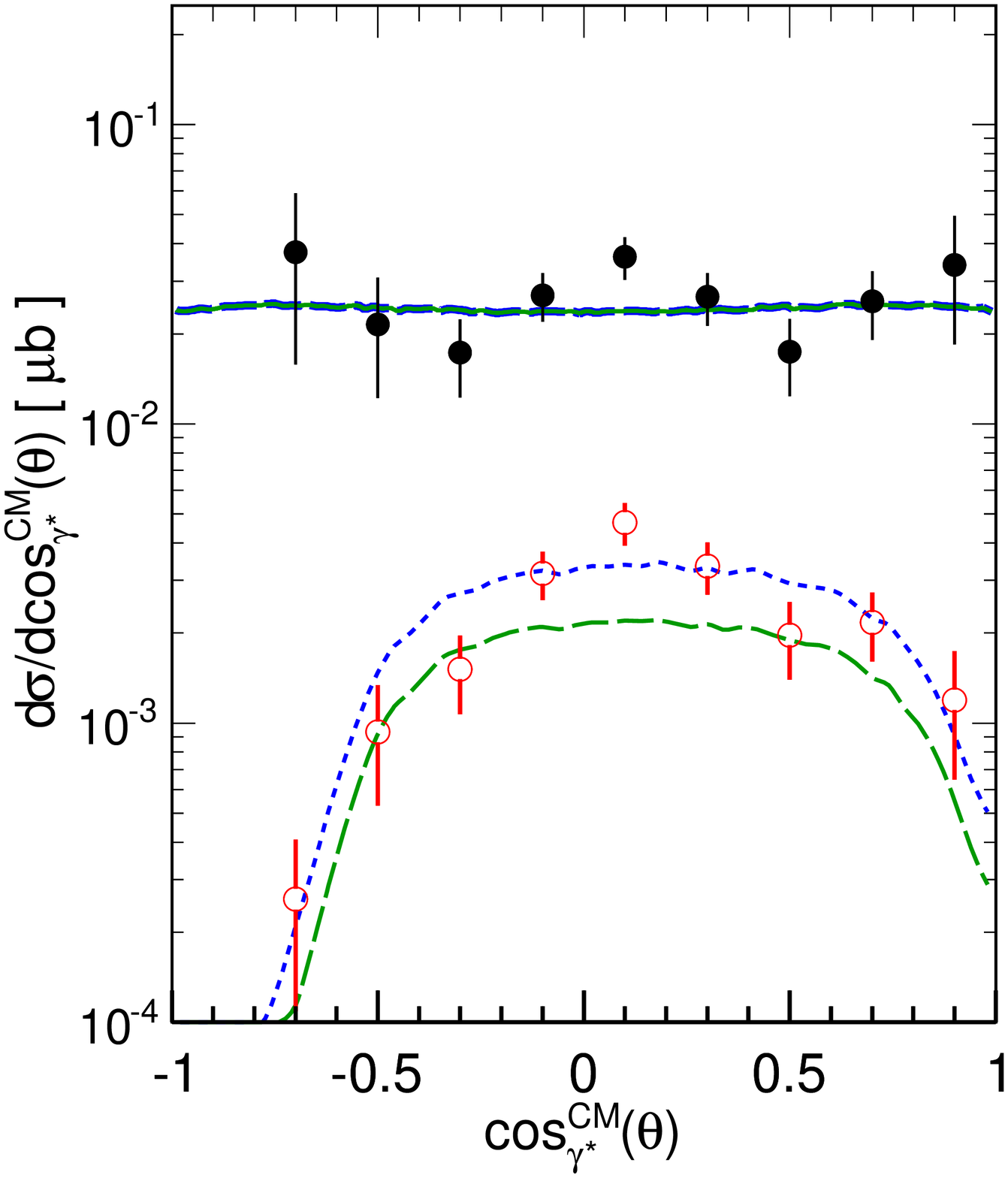}
\includegraphics{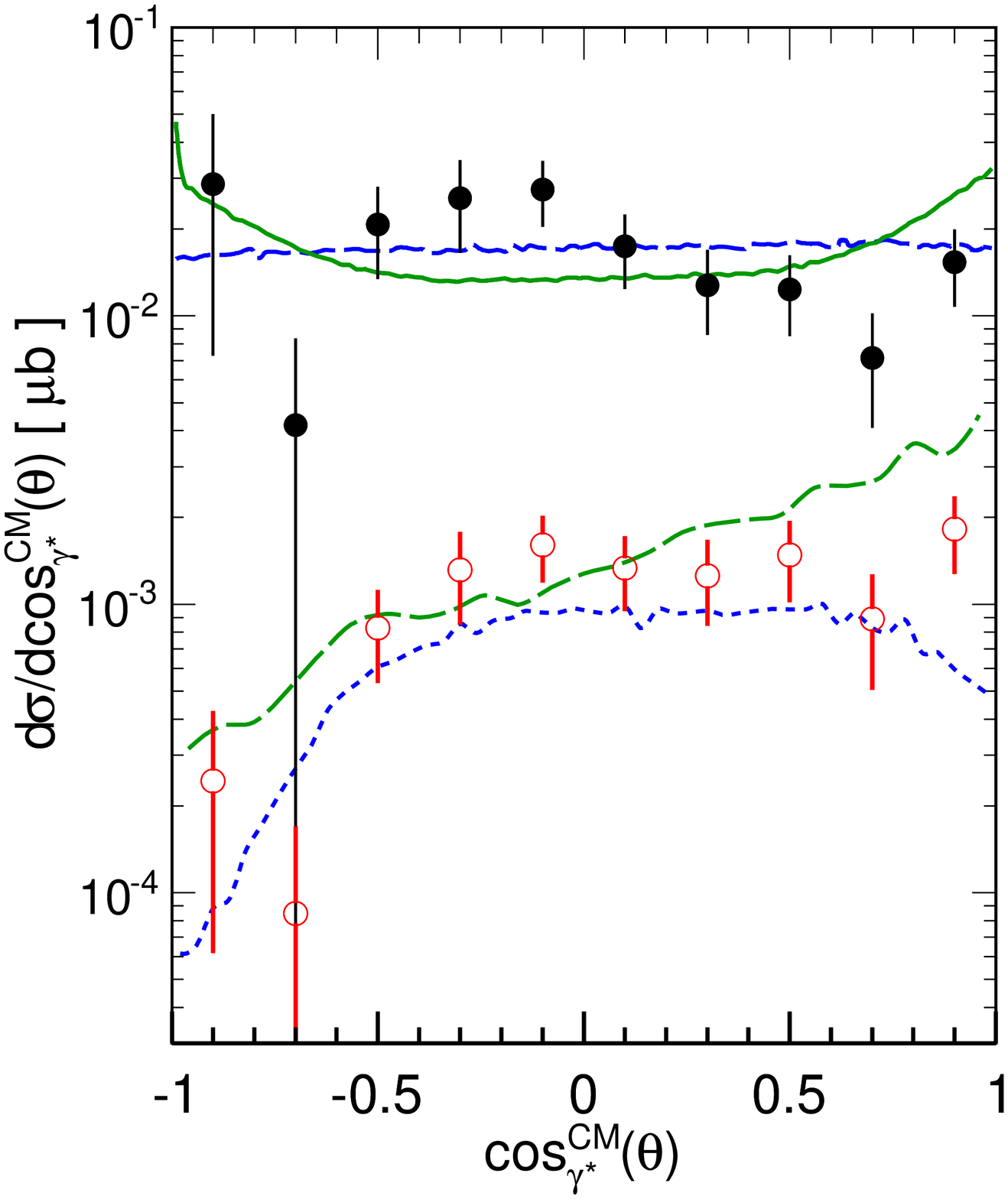}}

\caption{(Color online) Differential cross sections for the dielectron production in the $npe^+e^-$ channel as a function of the virtual photon emission angle in the c.m.s. for $0.14 < M_{e^+e^-} < 0.28$ GeV/c$^2$ (left panel) and $M_{e^+e^-} > 0.28$ GeV/c$^2$ (right panel). The red open dots present data within the HADES acceptance while the black full dots show the acceptance corrected data. The solid curves display predictions from the simulations in the full solid angle normalized to the experimental yield: the green curve represents model A (in the low-mass bin mostly $\Delta$, in the high-mass bin mostly $\rho$) and dashed blue represents the model B. The dashed/dotted curves show the respective distributions in the acceptance for model A (dashed green) and B (dotted blue) (see text for details).}
\label{cos_gam}
\end{figure}

Figure~\ref{cos_gam} presents similar angular distributions as discussed above but for the virtual photon. The distributions are also strongly biased by the HADES acceptance, which suppresses virtual photon emission in the forward and even more strongly in the backward direction. In the lower mass bin, where the $\Delta$ contribution is dominant, a deviation from the isotropic distribution could be expected due to the polarization of the $\Delta$ resonance. However, the experimental distributions are compatible with an isotropic emission, as assumed in the simulation. In the larger mass bin, it is interesting to see that the model A (solid green curve) predicts a significant anisotropy, related to the angular momentum in the double-$\Delta$ system for the $\rho$ emission by the charged pion line between the two $\Delta$s, which is the dominant contribution in this mass bin. However, our data present a different trend, which seems also to deviate from isotropy but with a smaller yield for the forward and the backward emission. Unfortunately, as already mentioned for the proton angular distributions, we cannot really verify these distributions based on the hypothesis of an emission by the charged pion between two nucleons, since the calculations in \cite{shyam2} do not provide them and the distribution of the model B remains here rather flat (dashed blue curve if Fig.~\ref{cos_gam}).

Finally we present distributions of leptons in the rest frame of the virtual photon. These observables are predicted to be particularly sensitive to the time-like electromagnetic structure of the transitions \cite{titov}. Indeed, for the Dalitz decay of the pseudo-scalar particle, like  pion or $\eta$ mesons, the angular distribution of the electron (or positron) with respect to the direction of the virtual photon in the meson rest-frame is predicted to be proportional to $1+cos^2(\theta_e)$. These predictions were confirmed in our measurements of the exclusive pion and eta meson decays in proton-proton reactions \cite{pp2GeV}. 

For the $\Delta$ Dalitz decay, the angular distribution has a stronger dependence on the electromagnetic form factors due to the wider range in $e^+e^-$ invariant masses. Assuming the dominance of the magnetic transition in the $\Delta \rightarrow Ne^+e^-$ process, the authors of \cite{titov} arrive at the same distribution as for the pseudo-scalar mesons. Concerning the elastic bremsstrahlung process, only predictions based on the soft photon approximation exist in the literature \cite{titov}. According to this model, the corresponding angular distributions show at our energies a small anisotropy with some dependence on the dielectron invariant mass. On the other hand, the angular distribution of leptons from the $\rho$-meson decay from pion annihilation, measured with respect to the direction of the pion in the virtual photon rest frame, has a strong anisotropy, i.e. $\propto 1-\cos^2(\theta_e)$.

\begin{figure}
\hspace{0mm}
\resizebox{0.48\textwidth}{0.25\textheight}{
\includegraphics{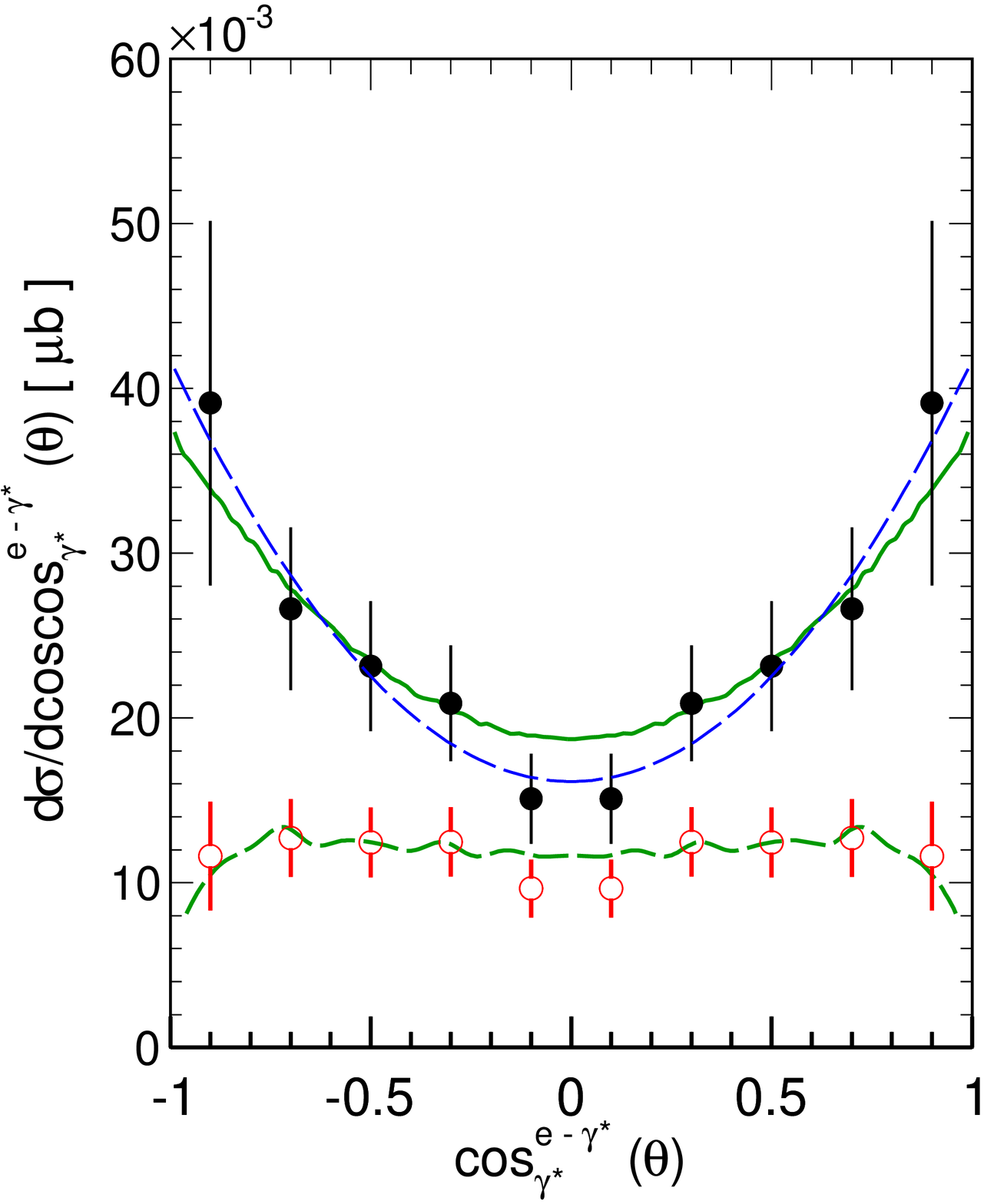}
\includegraphics{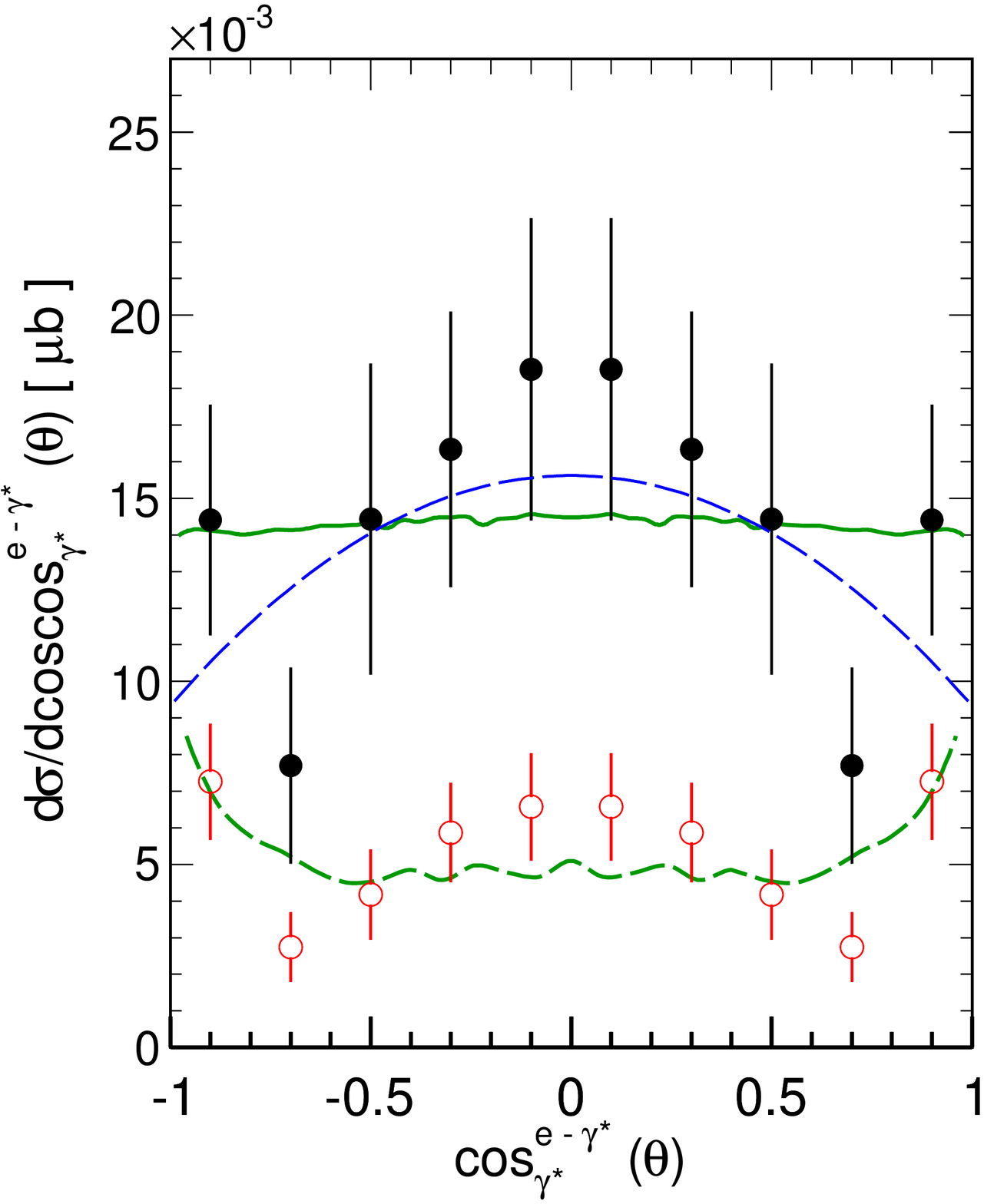}}

\caption{(Color online) Angular distributions of the leptons in the rest frame of the virtual photon, calculated in the $pe^+e^-$ rest frame and for the dielectron mass of $0.14<M_{e^+e^-}<0.28$ GeV/c$^2$ (left panel), and with respect to the direction of the charged pion exchange for dielectrons with $M_{e^+e^-}>0.28$ GeV/c$^2$ (right panel). The open red dots present the data within the HADES acceptance (multiplied by a factor 5), while the full black dots show the acceptance corrected data. The solid green  curves display predictions from the simulations in the full solid angle for model A, where the dominant source is the $\Delta$ (low-mass bin) and the $\rho$ (high-mass bin). The dashed green curve represents the data within the HADES acceptance, normalized to the experimental distributions. The dashed blue curve shows a fit with a function $A (1+B\cos^2(\theta_e))$. }

\label{helicity}       
\end{figure}

Figure~\ref{helicity} presents the respective $e^+$ and $e^-$ angular distributions for the experimental data and the two bins of the dielectron invariant mass. The distributions are symmetric due to the fact that both angles, between electron and $\gamma^*$ as well as positron and $\gamma^*$, in the rest frame of the virtual photon, have been plotted. For the left panel (bin with the smaller masses, $0.14<M_{e^+e^-}<0.28$ GeV/c$^2$) the distribution has been calculated with respect to the $\gamma^*$ direction, obtained in the $pe^+e^-$ rest frame, while for the right panel (bin with the larger masses, $M_{e^+e^-}>0.28$ GeV/c$^2$) it has been calculated with respect to the direction of the exchanged charged pion momentum. The latter one has been calculated as the direction of the vector constructed from the difference between the vectors of the incident proton and reconstructed emitted neutron and boosted to the rest frame of the virtual photon. The open red symbols present the data within the HADES acceptance while the full black symbols show the acceptance corrected data. The solid green curve displays a prediction from the simulation in the full solid angle while the dashed green curve, is normalized to the experimental distributions within the HADES acceptance for a better comparison of the shape. The dashed blue curve shows a fit with a function $A (1+B\cos^2(\theta_e))$. In the lower mass bin the data follow the distribution expected for the $\Delta$, $B=1.58 \pm 0.52$ and the fit almost overlays with the simulated distribution. This seems to confirm the dominance of the $\Delta$ in this mass bin, in agreement with both models. However, it would be interesting to test the possible distortion that could arise due to contribution of nucleon graphs, following \cite{shyam2}. For these graphs, the distribution of the $e^+$ or $e^-$ angle in the virtual photon rest frame should depend on the electric and magnetic nucleon form factors in a very similar way to the $e^+e^- \leftrightarrow \bar{p} p$ reactions, i.e. following $|G_M|^2 (1+\cos^2\theta)+(4m_p^2/q^2)|G_E|^2\sin^2\theta$, where $m_p$ is the proton mass. In the calculation of \cite{shyam2}, the anisotropy of the $e^+$ ($e^-$) angular distribution should therefore derive from the VDM form factor model. A similar fit to the higher mass bin in the same reference frame (not shown) gives a significantly smaller anisotropy $B=0.25 \pm 0.35$ which changes the sign, when the distribution of the lepton with respect to the exchanged charged pion is fitted ($B=-0.4 \pm 0.20$), as shown in Fig.~\ref{helicity} (right panel). The latter may indicate the dominance of the $\rho$ decay, as suggested by both models \cite{shyam2,clement}.

 \section{Summary and Outlook}
 \label{summary_outlook}
We have shown results for the quasi-free exclusive $np\to np e^+ e^-$ channel measured with HADES using a deuterium beam with a kinetic energy $T=1.25$ GeV/nucleon. The $e^+e^-$ invariant-mass differential cross section presents a similar excess with respect to the one measured in the $pp\to ppe^+e^-$ channel as previously observed for the corresponding inclusive $e^+e^-$ distributions, hence confirming the baryonic origin of this effect. In addition, the detection of the proton provides additional observables (invariant masses, angular distributions) which bring strong constraints for the interpretation of the underlying process. We tested two models which provided an improved description of the inclusive $e^+e^-$ production in the $n-p$ reaction at large invariant masses. The first one consists of an incoherent cocktail of dielectron sources including (in addition to $\pi^0$, $\Delta$ and $\eta$ Dalitz decay) a contribution from the $\rho$-like emission via the double-$\Delta$ excitation following the suggestion by Bashkanov and Clement \cite{clement}. The second model is based on the Lagrangian approach by Shyam and Mosel \cite{shyam2} and provides a coherent calculation of the $np\to npe^+e^-$ reaction including nucleon and resonant graphs. In both models, the enhancement at large invariant masses is due to the VDM electromagnetic form factor which is introduced for the production of the $e^+e^-$ pair from the exchanged pion. The evolution of the shape of the experimental $e^+$ and $e^-$ angular distribution in the $\gamma^*$ rest frame seems to confirm the emission via an intermediate virtual $\rho$ at the largest invariant masses. Since this process is absent in the reaction $pp\to ppe^+e^-$, it provides a natural explanation for the observed excess.

The different nature of the graphs at the origin of this $\rho$-like contribution in the two models is reflected in the invariant mass distributions. A better description of the experimental distributions is obtained with the model B, where the effect is related to the nucleon charge-exchange graphs. However, this conclusion should be tempered by the fact that we had to introduce a hypothesis for the angular distributions of the final products, which were not provided by the models. The agreement is also not perfect, which points to missing contributions. On the other hand, it is clear that the double-$\Delta$ excitation process is expected to play a role in the $e^+e^-$ production. In \cite{clement}, the corresponding amplitude is deduced from the modified Valencia model, which gave a fair description of $np\to np\pi^+\pi^-$ measured by HADES at the same energy \cite{hades_2pi}. A realistic test of the contribution of the double-$\Delta$ excitation to the $e^+e^-$ production can be only supplied once the effect is included as a coherent contribution in a full model including the nucleon and $\Delta$(1232) graphs, like the OBE calculation \cite{clement} and if all distributions are provided for a comparison with the differential distributions measured in the exclusive $np\to npe^+e^-$. The present analysis should serve as a motivation for such a complete calculation.

The first observation of an unexplained  dielectron excess measured in the inclusive $n-p$ reaction with respect to the $p-p$ reaction triggered a lot of theoretical activity and raised interesting suggestions of mechanisms specific to the $n-p$ reaction. Understanding in detail the $e^+e^-$ production in $n-p$ collisions is a necessary step towards the description of $e^+e^-$ production in heavy-ion collisions where medium effects are investigated. On the other hand, the description of the $np\to npe^+e^-$ process is challenging because it implies many diagrams with unknown elastic and transition electromagnetic form factors of baryons in the time-like region. We have shown that our exclusive measurement of the quasi-free $np\to npe^+e^-$ reaction at $T=1.25$ GeV is sensitive to the various underlying mechanisms and in particular sheds more light on contributions which are specific to the $n-p$ reaction. While definite conclusions can only be drawn when more detailed calculations are available, we also expect additional experimental constraints from the on-going analysis of the $np\to de^+e^-$ reaction, also measured by HADES at the same energy.

\section{Acknowledgements}

The HADES Collaboration gratefully acknowledges the support by the grants LIP Coimbra, Coimbra (Portugal) PTDC/FIS/113339/2009, UJ Kraków (Poland) NCN 2013/10/M/ST2/00042, TU M\"unchen, Garching (Germany) MLL M\"unchen: DFG EClust 153, VH-NG-330 BMBF 06MT9156 TP5 GSI TMKrue 1012 NPI AS CR, Rez, Rez (Czech Republic) GACR 13-06759S, NPI AS CR, Rez, USC - S. de Compostela, Santiago de Compostela (Spain) CPAN:CSD2007-00042, Goethe University, Frankfurt (Germany): HA216/EMMI HIC for FAIR (LOEWE) BMBF:06FY9100I GSI F\&E, IN2P3/CNRS (France). The work of A.V. Sarantsev is supported by the RSF grant 16-12-10267.

\end{document}